\documentclass[cits]{PoS}

\usepackage{graphicx}
\usepackage{epstopdf}

\def\a{\alpha}
\def\b{\beta}

\def\D{\Delta}

\long \def \blockcomment #1\endcomment{}

\newcommand{\bee}{\begin{equation}}
\newcommand{\ee}{\end{equation}}
\newcommand{\beea}{\begin{eqnarray}}
\newcommand{\eea}{\end{eqnarray}}

\title{Scaling properties of many-fermion systems from MCRG studies}

\ShortTitle{Scaling properties of many-fermion systems from MCRG studies}

\author{\speaker{Anna Hasenfratz}%
       \\
       Department of Physics, University of Colorado, Boulder Colorado 80303 USA\\
       E-mail: \email{anna@eotvos.colorado.edu}}

\abstract{Monte Carlo renormalization group methods were designed to study the phase structure and critical behavior of statistical systems. They are well suited to determine the running coupling and to investigate the properties of  fixed points of gauge-fermion  models, including the existence of conformal infrared fixed points in many-fermion systems. I discuss the implementation of the 2-lattice matching method and present results for SU(3) gauge theories with $N_f=0,12$ and 16 fundamental fermion flavors.
}

\FullConference{The XXVII International Symposium on Lattice Field Theory\\
                July 26-31, 2009\\
                Peking University, Beijing, China}

\begin{document}

\section{Introduction}

 SU(N) gauge theories with different number of  fermions in various representations can play an important role in Beyond-Standard-Model phenomenology.  It is important to clearly 
 establish if a given model is QCD-like, governed by its perturbative fixed point,  or conformal with an infrared fixed point (IRFP) at a finite gauge coupling. 
  Identifying the renormalization group (RG) structure  is not particularly difficult far away from the bottom of the conformal window. Unfortunately the most interesting case, distinguishing a confining, possibly walking model from a newly 
 emerging IRFP,  is much more difficult. 
 Such an IRFP likely occurs at strong coupling, requiring non-perturbative techniques, and lattice simulations offer the best method to attack the problem. 
 Since lattice  gauge models are always confining in the extreme strong coupling, identifying a confining, chirally broken  phase does not guarantee QCD-like behavior in the continuum limit.
 It is necessary to start in the weak coupling  where direct comparison with perturbation theory is possible, and connect it to intermediate/strong coupling.  If an IRFP is identified in the process, one can conclude that the theory has a conformal phase. If no IRFP is found, one has to show that the model exhibits confinement and chiral symmetry breaking. This has to be done  before the coupling is pushed into extreme strong coupling where universality is lost.  While there are well established lattice methods to carry out the above program,  the calculations are not easy and the emerging systematical and statistical errors are frequently much larger than  comparable QCD calculations.   Just as in QCD, it is essential to use  different methods and consider different lattice discretizations to control the errors when one attempts to distinguish  walking  and conformal theories and  predict their critical behavior. 
 
 In this work I describe  the application of the 2-lattice matching Monte Carlo renormalization group (MCRG) method to calculate the bare step scaling function and connect the weak and strong coupling regions\cite{Hasenfratz:2009ea}. This approach has the advantage to the frequently used Schroedinger functional (SF) method \cite{Luscher:1992zx,Luscher:1993gh,Appelquist:2007hu,Appelquist:2009ty,Hietanen:2009az,Shamir:2008pb,Bursa:2009tj} that it has a free  parameter that can be tuned for optimization and it uses several observables that provide  consistency checks. I apply the method to SU(3) gauge models with various number of fundamental fermions. In the $N_f=0$ pure gauge case I verify that MCRG gives consistent results compared to other methods. The $N_f=16$ flavor system clearly shows the existence of an IRFP with scaling dimension that is consistent with the perturbative one. In the $N_f=12$ model I am unable to draw a definite conclusion. The calculation has not identified an IRFP and I have not been able to push far enough into the strong coupling  to establish a confining phase  connected to the weak coupling region either. $N_f=12$ is very close to the confining-conformal boundary and will require a more careful investigation, possibly with a different action or renormalization group transformation.  

\section{The 2-lattice matching MCRG method}

The 2-lattice matching MCRG method  identifies pairs of couplings $(\b,\b^\prime)$  corresponding to a factor of $s$ change in the lattice spacing, $a(\b)= a(\b^\prime)/s$.  A sequence of these pairs can be used to predict the bare step scaling function and consequently the running coupling in theories governed by one relevant operator. 

MCRG methods explore the phases and fixed point structure of lattice systems using a  real space RG block transformation. The block variables are 
 defined as a  local average of
the original lattice variables and  a scale $s$ transformation decreases the lattice size by the factor $s$. 
 By integrating out the original
variables while keeping the block variables fixed, one removes the
ultraviolet fluctuations below the length scale $sa$.  The RG transformation does not change the  physical correlation length, but the lattice correlation 
length after $n$ blocking steps is $\xi^{(n)} =  s^{-n}\xi^{(0)}$. 
 The RG can have fixed points only when $\xi=\infty$ (critical) or
$\xi=0$ (trivial).
Flow lines starting near
the $\xi=\infty$ critical surface approach the fixed point in the irrelevant directions
but flow away in the relevant one. After a few RG steps the irrelevant
operators die out and the flow follows the unique renormalized trajectory (RT), independent of the original couplings. If the flow lines from
a pair of couplings $(\b,\b^{\prime})$ 
end up at the same point along the RT  but one requires one less 
step to do so,  the lattice spacings at $\b$ and $\b^{\prime}$  must
differ by a factor of $s$.  
In order to identify a pair of couplings $(\b,\b^{\prime})$ with $\xi'=\xi/s$
we have to show that after $n$ and $(n-1)$ blocking steps their blocked
actions are identical. Fortunately one does not need to calculate the blocked actions, it is sufficient to show that the expectation values of
every operator measured on configurations generated with one or the
other action are identical. Furthermore it is possible to create a configuration
ensemble with Boltzman weight of an RG blocked action by generating
an ensemble with the original action and blocking the configurations
themselves \cite{Swendsen:1979gn}.
This suggests the following procedure for the 2-lattice
matching:
\begin{enumerate}
\item Generate a configuration ensemble of size $L^{d}$ with action $S(\b)$.
Block each configuration $n$ times and measure a set of expectation
values on the resulting $(L/s^{n})^{d}$ set.
\item Generate configurations of size $(L/s)^{d}$ with action $S(\b^{\prime})$,
where $\b^{\prime}$ is  a trial coupling. Block each configuration $n-1$
times and measure the same expectation values on the resulting $(L/s^{n})^{d}$
set. Compare the results with that obtained in step 1. and tune the
coupling $\b^{\prime}$ such that the expectation values agree.
\end{enumerate}

There is considerable freedom in defining block variables. In this work I use a scale $s=2$ transformation
 \begin{equation}
V_{n,\mu}={\rm Proj[}(1-\alpha)U_{n,\mu}U_{n+\mu,\mu}+\frac{\alpha}{6}\sum_{\nu\ne\mu}U_{n,\nu}U_{n+\nu,\mu}U_{n+\mu+\nu,\mu}U_{n+2\mu,\nu}^{\dagger}]\,,\label{eq:block-trans}
\end{equation}
where ${\rm Proj}$ indicates projection to $SU(3)$.  
The role of the parameter $\alpha$ is to optimize the block transformation.
While the critical surface of a system is well defined, the location
of the fixed point itself is not physical, it can be changed by changing
the RG transformation. It is important to optimize the blocking so
its FP and RT  can be reached in a few RG steps. The optimal blocking is characterized
by 
\begin{enumerate}
\item Consistent matching between the different operators: along the RT  all
expectation values should agree on the matched configuration sets. Any
deviation is a measure that the RT has not been reached.
\item Consecutive blocking steps should give the same matching condition.
\end{enumerate}

\section{Simulation results}
The following numerical results were obtained with plaquette gauge action and nHYP smeared \cite{Hasenfratz:2001hp,Hasenfratz:2007rf} staggered fermions.  I have generated 2-300 configurations separated by 10 molecular dynamics steps at each coupling value. The blocking transformation is given in Eq. \ref{eq:block-trans} and I have used 3 operators in the matching \cite{Hasenfratz:2009ea}. 
\subsection{$N_f=0$ pure gauge SU(3) gauge model}
 The  bare step scaling function  of the pure gauge SU(3) model was studied in Refs.\cite{Bowler:1984hv}
with the 2-lattice matching method.  I have  repeated some of those
calculations with a different block transformation and extended them
to larger volumes and statistics. Where they overlap, the results I present below are consistent with the original calculations. 
This section mainly serves as a test of the method.

\begin{figure}
\includegraphics[scale=0.8]{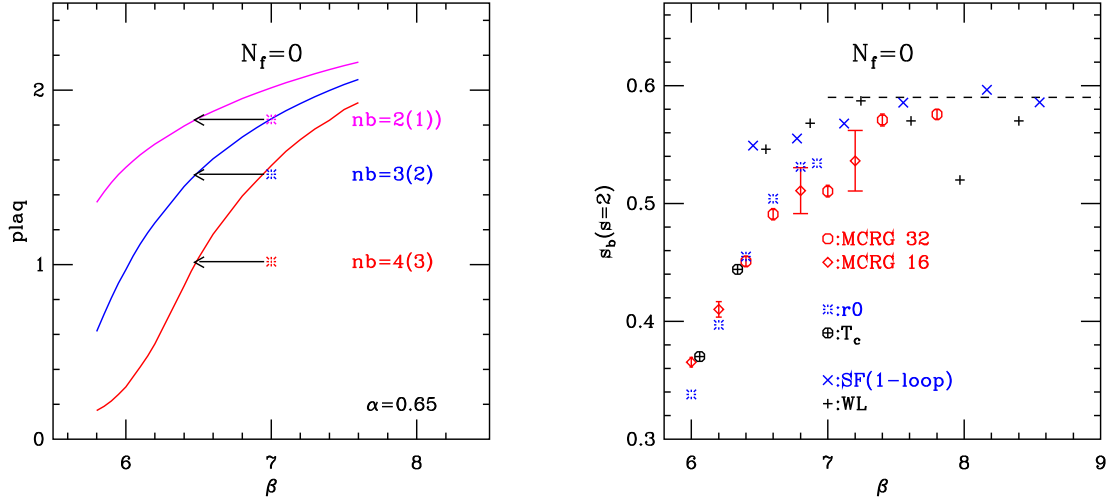}
\caption{Left panel: The matching of the plaquette for pure gauge SU(3) theory. Right panel: The bare step scaling function $s_b(\b;s=2)$ for the pure gauge SU(3) system
as predicted by different methods.  \label{fig:fig1}}
\end{figure}

Figure \ref{fig:fig1}  illustrates the 2-lattice MCRG.
The left panel  shows the matching of the plaquette 
 with  the blocking parameter   $\a=0.65$ RG transformation.
 $32^{4}$ volume simulations
were done  at $\beta=7.0$, and the magenta, blue and red
symbols give the values of the blocked plaquette after 2, 3 and 4 
blocking steps. The solid curves interpolate the plaquette values, measured at many couplings on $16^4$ volumes,  after
1, 2 and 3 blocking steps. The $32^{4}$
data match the $16^{4}$ values at $\beta^{\prime}=6.49$  for all blocking levels.
 The final blocked volume is   $2^{4}$, but finite size
effects are minimal as one always compares observables on the same
volume.   

The matching can be repeated with different operators and RG transformations. 
The difference between the matched couplings,
$\D\b = \b -\b^{\prime}$,
becomes less and less dependent on the blocking parameter as the blocking level increases. The step scaling function is defined at $n_b=\infty$ as 
$s_b(\b) = \rm{lim}_{n_b\to\infty} \D\b$. At finite $n_b$ I approximate  $s_b(\b)$  as $\D\b$ at the optimal blocking parameter, defined as where $\D\b$ does not change during the last two blocking steps. This definition clearly has systematical errors that can be reduced or eliminated only if the calculation is repeated on larger volumes.

In the right panel of Figure \ref{fig:fig1} I compare the MCRG results for the bare step scaling function with predictions from
  other methods, including  the 1-loop perturbative prediction, $s_b^{(\rm{pert})}(s=2)=0.59$ (dashed line).
  Note that I show errors only for the MCRG results. Errors from the $16^4 \to 8^4$ matching are considerably larger than from $32^4 \to 16^4$; this is mainly due to systematic effects. 
  As for the other determinations, I consider the  Sommer
parameter $r_0$ \cite{Sommer:1993ce,Necco:2001xg}, the critical temperature $T_c$ \cite{Boyd:1996bx}, as well as the SF calculation  \cite{Luscher:1993gh} and a  recently proposed
alternative based on Wilson loop matching (WL) \cite{Bilgici:2009kh} for the running coupling.

In the scaling regime the different approaches should give the same prediction for $s_b(\b)$. It is very satisfying to see the agreement between
 MCRG, $r_0$ and $T_c$ even at relatively strong couplings. In the range $\beta \in (6.0,7.0)$ the predicted
values differ considerably from the 2-loop perturbative results, but for $\beta\ge7.0$ 
 both
 the  SF, WL and MCRG  methods approach $s_b^{(\rm{pert})}$.  The 2-lattice MCRG matching method is competitive with other methods in determining the running coupling of asymptotically free theories.

\subsection{$N_f=16$ flavor SU(3) model}
I have used the procedure outlined in the previous section to study the 16-flavor SU(3) model.   
Perturbation theory predicts the existence of an IRFP in the chiral limit  at a fairly weak gauge coupling.
Numerical simulations, based on the SF and WL ratio approaches, have verified the backward (towards weak coupling) flow of the gauge coupling \cite{Heller:1997vh,Fodor:2009wk}.
Calculations to identify the IRFP are cleanest in the massless limit, but in practice  it is easier to introduce a small mass in the simulations. I used  $m=0.01-0.02$ but verified that none of the observables had any measurable dependence on the mass and for all practical purposes the simulations can be considered as in the chiral limit. I have generated  $16^4$ and $8^4$ lattices at various values of the gauge coupling to map the bare step scaling function.

\begin{figure}
\begin{center}
\includegraphics[scale=0.7]{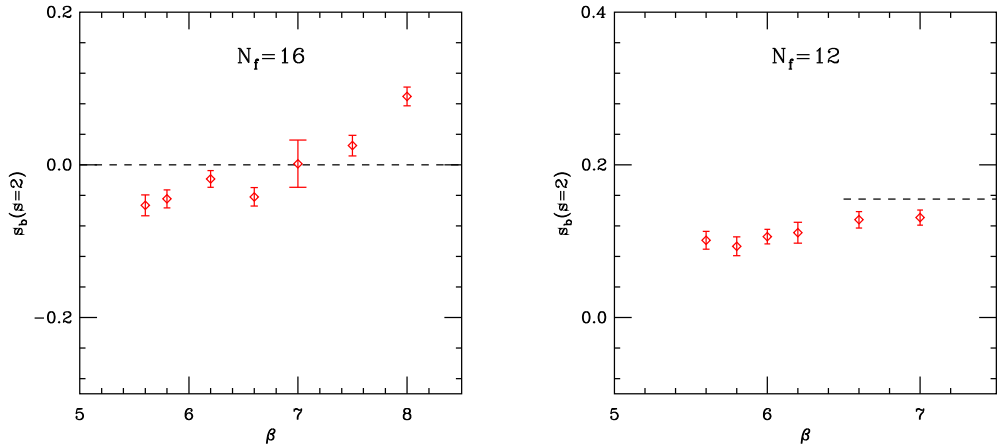}
\end{center}

\caption{The bare step scaling function $s_b(\b;s=2)$. Left panel:  $N_f=16$; right panel: $N_f=12$.}

\label{fig:fig2}

\end{figure}

The 2-lattice matching method picks up the relevant operator of a FP. However at the IRFP of the 16 flavor model there is no relevant operator in the massless limit. As $n_b\to\infty$ matching looses its meaning completely since all the flow lines converge to the fixed point. With finite $n_b$ one is likely to pick up the flow in the least irrelevant direction. Perturbation theory predicts that the critical index governing the flow  of the gauge coupling at the IRFP  is very small for  $N_f=16$. The gauge coupling is nearly marginal and one expects the matching to follow this coupling.
The left panel of Figure \ref{fig:fig2} summarizes the results. The step scaling function is negative for $\b <7.0$ signaling the existence of an IRFP. At $\b=7.5$ and 8.0 $s_b(\b)>0$, so this block transformation has an IRFP somewhere around $\b\approx 7.0$. Note that  the results presented here are  different though consistent with Ref\cite{Hasenfratz:2009ea} as I have extended the statistics and improved the analysis since the publication. 
Fitting $s_b(\b)$ with a linear function  in the range $\b\in(5.8,7.5)$ predicts the slope 0.036(12). This is somewhat larger but not inconsistent with the perturbative value of 0.022. 

\subsection{$N_f=12$  flavor SU(3) model}

The $N_f=12$ system is very close to the opening of the conformal phase. Calculations with SF using unimproved staggered fermions identified an IRFP, but spectral measurements might favor a QCD-like interpretation \cite{Appelquist:2007hu,Appelquist:2009ty,Fodor:2009wk,Jin:2009mc,Bilgici:2009nm}. I have used the 2-lattice matching MCRG to connect the perturbative weak coupling phase to the strong coupling, looking for an IRFP or the
emergence of a confining, chirally broken phase. The calculation itself is very similar to the $N_f=16$ case. I used $16^4$ and   $8^4$ lattices and simulated nHYP staggered fermions at $m=0.01$.

The results for the bare step scaling function is shown on the right panel of Figure \ref{fig:fig2}. The dashed line indicates the perturbative $s_b^{\rm{pert}}(s=2)=0.155$ value. The numerical results approach this for large gauge couplings and show a slight decrease  at smaller values. The last data point is at $\b=5.6$. For $\b<5.6$  matching with the block transformation of Eq. \ref{eq:block-trans} on $16^4 \to 8^4$ lattices does not give consistent results for the different operators. This does not mean that matching is not possible, only that this RG transformation does not reach the FP or its renormalized trajectory in 3/2 blocking steps. 

The MCRG results indicate that there is no IRFP in the $N_f=12$ system for $\b>5.6$. $16^4$ volumes are deconfined at this coupling, so this information is not sufficient to determine if the model is conformal or QCD-like. If the system is conformal, here has to be a bulk phase transition separating the conformal phase from the confining strong coupling lattice phase at least in the chiral limit.  I have  studied the coupling range from $\b=5.6$   to $\b=2.5$, monitoring the plaquette, blocked plaquettes, and several other small Wilson loops. None of these observables showed sign of a  phase transition. $16^4$ lattices remained deconfined for $\b\ge4.2$ (the smallest coupling I performed $16^4$ simulations) and $8^3\times16$ lattices remained deconfined  for $\b\ge2.5$. All I can say at this point is that if the $N_f=12$ flavor system with nHYP staggered fermions  is conformal, its bulk transition is at $\b<2.5$. If it is QCD-like, its lattice spacing at $\b=2.5$ is small enough to deconfine an $8^3 \times 16$ volume. On the other hand the plaquette at $\b=2.5$ is around 0.5 (out of 3): this is such a small value that smearing becomes problematic, and lattice artifacts likely overwhelm the system.

It is possible that a different RG transformation has an IRFP at an accessible coupling, or a different action, possibly with less efficient smearing, allows the identification of either an IRFP or direct connection to a confining phase.

\section{Summary}

Identifying the renormalization group structure of many-fermion theories is a difficult task in models that are near the confining - conformal boundary.  In this work I presented a method based on Monte Carlo renormalization group that can be used to effectively  calculate the bare step scaling function and connect the weak and strongly coupled regions. My results for the $N_f=0$ pure gauge system agree with other determinations for the step scaling function. For the $N_f=16$ model they show the emergence of an IRFP and give the critical exponent   of the gauge coupling consistent with the perturbative value. The $N_f=12$ case is much less clear. With MCRG I could not push into strong enough coupling to find a confining phase, neither did the calculation show an IRFP. Different block transformations or actions could overcome the problems and should be investigated in the future.

\section{Acknowledgment }

I have benefited from many discussions during the "Lattice 2009 Conference" in Beijing and the  "Strong Dynamics Workshop"  at the Lorentz Center, Leiden, NL.  The computations  of this project were carried out on the High Energy Physics cluster at the University of Colorado and on the USQCD  cluster at Fermilab.
This research was partially supported by the US Dept. of Energy and by the Project of Knowledge Innovation Program (PKIP) of Chinese Academy of Sciences, Grant No. KJCX2.YW.W10.

{\renewcommand{\baselinestretch}{0.86}
 \bibliography{lattice}
 \bibliographystyle{JHEP-2}}

\end{document}